\documentclass[useAMS,usenatbib]{mn2e}
\usepackage[dvips]{graphicx}
\usepackage{amssymb}

\newcommand{\eg}{{e.g.\ }}

\newcommand{\ie}{{i.e.\ }}

\def\aj{AJ}%
\def\apj{ApJ}%
\def\apjs{ApJS}%
\def\aap{A\&A}%
\def\aaps{A\&AS}%
\def\mnras{MNRAS}%
\title[Faint red halo around an edge-on disc in the UDF]{A faint red stellar 
halo around an edge-on disc galaxy in the Hubble Ultra Deep Field}
\author[S. Zibetti and A. M. N. Ferguson]{Stefano Zibetti$^{1}$\thanks{E-mail:
zibetti@MPA-Garching.MPG.DE} and Annette  M. N. Ferguson$^{1}$\\
$^{1}$Max-Planck-Institut f\"ur Astrophysik, Karl-Schwarzschild-Str. 1, 
D-85748 Garching bei M\"unchen, Germany}
\begin{document}

\date{Accepted . Received ; in original form 2004 April 23}

\pagerange{\pageref{firstpage}--\pageref{lastpage}} \pubyear{2004}

\maketitle

  \label{firstpage}

\begin{abstract}
We analyse the detailed structure of a highly-inclined
($i\ga~80\degr$) disc galaxy which lies within the Hubble Ultra Deep
Field (UDF). The unprecedented depth of the UDF data allow disc and
extraplanar emission to be reliably traced to surface brightness
levels of $\mu_{V,i,z}\sim29-30$ mag arcsec$^{-2}$ (corresponding to
rest-frame equivalents of $\mu_{g,r,i}\sim28-29$ mag~arcsec$^{-2}$) in
this redshift $z=0.32$ system.  We detect excess emission above the
disc which is characterised by a moderately-flattened
($b/a\sim0.6$) power-law ($I \propto R^{-2.6}$).  The
structure and colour of this component are very similar to the stellar
halo detected in an SDSS stacking analysis of local disc
galaxies \citep{zib04} and lend support to the idea that we have
detected a stellar halo in this distant system.  Although the peculiar 
colours of the halo are difficult to understand in terms of normal
stellar populations, the consistency found between the UDF and SDSS
analyses suggests that they cannot be easily discounted.
\end{abstract}


\section{Introduction}
In the currently-favoured $\Lambda$CDM framework, galaxy formation
proceeds hierarchically with small structures forming first and later
merging and accreting to form large galaxies.  In these models, a
significant fraction the stars which reside in the stellar halo and
thick disc are tidally-stripped from small satellites as they fall
within the host potential \citep[\eg][]{abadi03}.  The structure,
composition and ubiquity of these faint stellar components are thus
expected to reflect the details of the galaxy assembly process
\citep[\eg][]{bull04}.\\
\indent Observations of stellar haloes and thick discs in external galaxies
are extremely challenging. These faint components typically have
surface brightnesses $\mu_{R}\gtrsim 28$~mag~arcsec$^{-2}$, which
corresponds to $\gtrsim 7$ magnitudes fainter than the
sky. Measurements of diffuse emission require flat-fielding and sky
subtraction uncertainties, as well as scattered light effects, to be
significantly less than $0.1$ per cent. Until recently, few
observational constraints were available with which to confront
models.  While there have been some detections of faint stellar
components in external galaxies
\citep[\eg][]{sack94,morr97,lequeux_etal98,abe99}, there
have also been non-detections \citep[\eg][]{zheng99,fry99}, leading one to
question to what extent the results reflect real variance as opposed
to systematic effects \citep[see][ hereafter Z04, for a full discussion]{zib04}.
The alternative technique of using wide-area resolved
star counts to study stellar haloes sidesteps the difficulties
inherent in quantifying extremely faint diffuse emission, but 
can only be applied to a handful of nearby galaxies \citep[\eg][]{ferg02}.\\
\indent Recently Z04 have conducted the first statistical study
of stellar halo emission by stacking $\gtrsim1000$ homogeneously
rescaled edge-on galaxies from the Sloan Digital Sky Survey (SDSS).
This technique has allowed quantitative analysis of the $``$mean''
stellar halo to $\mu_r\sim31$ mag~arcsec$^{-2}$.  These
authors find a halo characterised by a moderately-flattened
($b/a\sim0.6$) power-law ($I \propto R^{-2}$) and puzzling
colours ($g-r=0.65$, $r-i=0.6$) that cannot be easily explained
by normal stellar populations. The detection of flatter
red stellar envelopes around extreme late-type edge-on disc
galaxies by \cite{dal02} may represent the higher surface brightness
regions of these haloes.\\
\indent The recent release of the Hubble Ultra Deep Field (UDF), a public survey carried out with
the Advanced Camera for Surveys using Director's Discretionary time in
Cycle 12 \citep{beck03}, provides a rare opportunity to quantitatively study
galaxy structure to the depths where halo emission should prevail.
This {\it Letter} reports the detection of a faint red stellar
halo around a $z=0.32$ edge-on galaxy in the
UDF.
\section{Observations and Analysis}
The Hubble Ultra Deep Field (UDF) consists of 400 orbits of
integration on a single 11.3 arcmin$^2$ field lying within the Chandra
Deep Field South GOODS area, centred at RA$=03^{h}32^{m}39\fs{0}$,
Dec$=-27\degr47\arcmin29\farcs1$ (J2000).  Four filters have been
utilised: $B$(F435W), $V$(F606W) (each for 56 orbits, 37.5 hours) and
$i$(F775W), $z$(F850LP) (each for 144 orbits, 96.4 hours). The UDF
represents the deepest observations yet obtained with HST, or
any ground-based telescope.

\begin{figure*}
\includegraphics{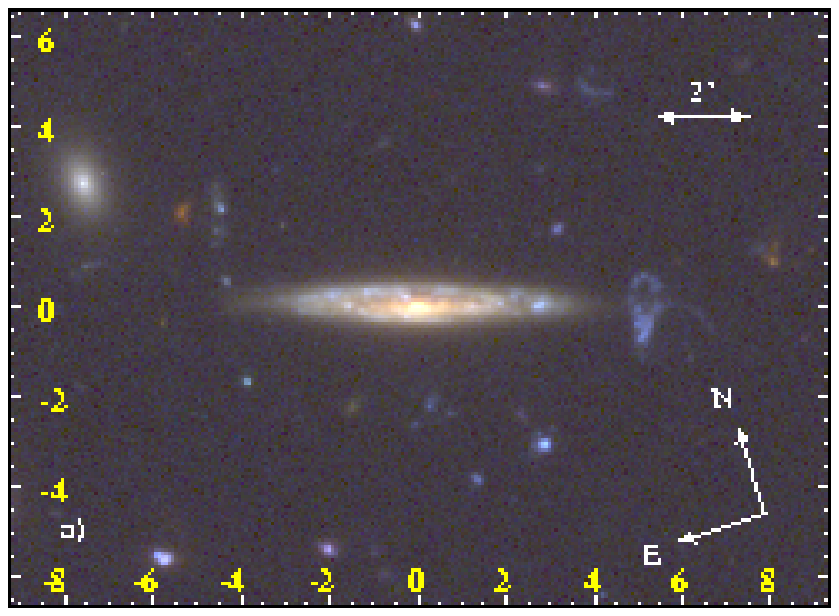}
\includegraphics{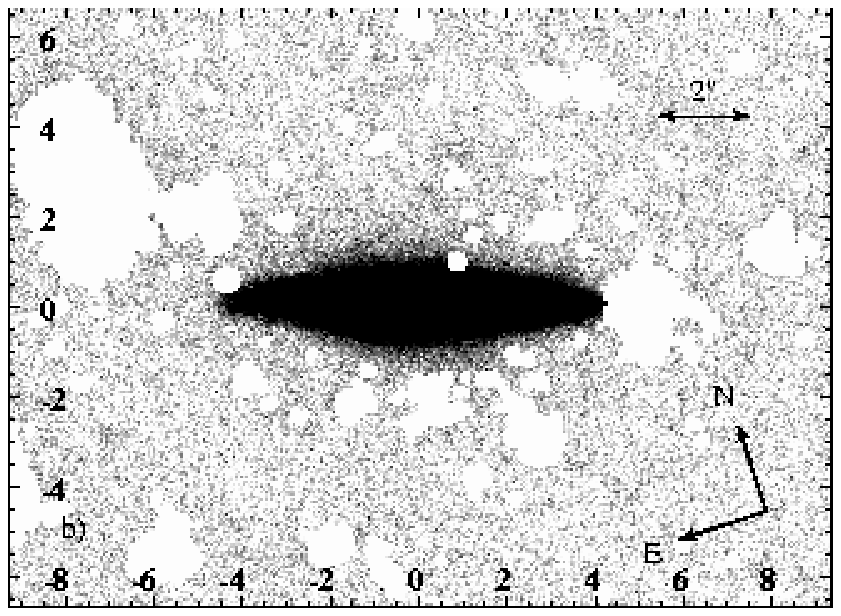}
\caption{a) Full resolution colour composite RGB ($(i+z)$-$V$-$B$) image of the galaxy 
showing whole dynamic range. b) $i+z$ band inverted gray-scale image,
with enhanced contrast in the low surface brightness levels and masked
regions superposed. Intensity scaling is a square root function in
both panels. Axes show the offset in arcseconds with respect to the
galaxy centre.  }
\label{picture}
\end{figure*}

We focus on a relatively large, well-formed disc galaxy located at
RA$=03^{h}32^{m}41\fs{1}$, Dec$=-27\degr48\arcmin52\farcs9$ (J2000).
This galaxy is clearly detected in all 4 bands and is identified with
source \#31611 at $z=0.322$ in the COMBO-17 photometric redshift
survey of \cite{wolf04}.  The high inclination of this system makes it
an ideal target for a study of faint extraplanar emission. At this
redshift, 1 arcsec=4.7~kpc and the lookback time is
3.6~Gyr\footnote{For $H_0$=70km sec$^{-1}$ Mpc$^{-1}$, $\Omega_0$=1,
$\Omega_\Lambda$=0.7}. We note the presence of a ring-like structure
of blue knots lying along the western major axis (see
Fig. \ref{picture}(a)); this structure has a photometric redshift of
unity according to \cite{wolf04} and thus appears as a chance
alignment.

Our analysis is based on the reduced UDF data v1.0 released
to the community by STScI on March 9, 2004\footnote{Available from
http://www.stsci.edu/hst/udf/release}.  The data were pre-processed
using the standard ACS/WFC pipeline and subsequently
combined using the Multidrizzle package.  The
observations were obtained using a 4-point dither pattern to provide
sub-pixel sampling, as well as a larger-scale pattern to
cover the 2 arcsec gap between the ACS/WFC chips. The final pixel
scale after drizzling is 0.6 of the original,
corresponding to 0.03 arcsec.  Note that all magnitudes are on the
{\sl AB}-system with zero-points provided by STScI.  We have
additionally applied a correction for Galactic extinction using  \cite{schlegeldust}.

Using the geometric parameters obtained from \emph{STSDAS-ellipse}
fitting of the $i$-band isophotes, we have extracted $1200\times1200$
pixel$^2$ frames in each band, centred on the galaxy and rotated
according to the average position angle of the isophotes at
$\mu_{i}\sim24$~mag arcsec$^{-2}$.  The RGB colour composite image of
the galaxy is shown Fig. \ref{picture} (a), where we use the $B$-band
image for the blue channel, $V$ for the green, and $i+z$ for the red.

We use SExtractor v2.3 \citep{sextractor} to extract very low
signal-to-noise segmentation images of the frames, adopting a
$5\times5$~pix$^2$ gaussian smoothing kernel ($\sigma=2$~ pixels), a
1-$\sigma$ detection threshold and a 25 pixel minimum detection area.
After excluding the segment(s) corresponding to the galaxy and
manually editing the mask to include sources that were not correctly
deblended from the galaxy, we combine the masks in the four bands and
``grow'' the resulting mask by 2 pixels, in order to have better
coverage of the extended diffuse sources.  The masked $i+z$-band image
is shown in Fig. \ref{picture} (b), with an enhanced contrast in order
to show the lowest intensity levels.

Visual inspection indicates that the galaxy is close to edge-on and
this is confirmed by the measured isophotal axial ratio of 0.2 at
$\mu_{i}\sim24$~mag arcsec$^{-2}$.  Assuming the standard formula for
an infinitely thin disc (cos~{\it i}=b/a), this corresponds to an
inclination of 78$\degr$.  For the more realistic case of finite
thickness, the inclination will be somewhat greater than this.
Fig. \ref{picture} a) also suggests a  late-type classification
for the galaxy based on prominent star-forming spiral arms, a
significant amount of dust in the inter-arm region and a very small
 bulge, if present at all.  In Fig. \ref{picture} b), one can clearly
see the presence of red extraplanar emission which will be analysed in
significant detail in the following section.\

Although the UDF science images are already background subtracted, we
estimate and subtract the residual average local background using an
annulus (inner radius$=400$ pixels or 56~kpc, width $=200$ pixels or
28~kpc) around the galaxy on the masked images . This annulus is
chosen to lie outside the region where we might reasonably expect
faint halo emission to be present.  The maximum correction corresponds
to a surface brightness of $\sim31$~mag arcsec$^{-2}$ (measured in the
z-band); comparison to the faintest measured z-band surface brightness
indicates that the level of uncertainty due to the background is no
more than 20\%.  Smaller background uncertainties are found in all
other bands.

The primary photometric analysis is performed on the masked images
using an in-house developed C code which provides us with the average
pixel value as a function of radius within a wedge-shaped
region with apex located at the galaxy centre. As we believe the galaxy is close to
edge-on, this technique provides an alternative to making simple major
and minor axis cuts and is particularly well-suited to the study of very faint emission
at large radius since relatively more pixels are averaged in these
parts. Errors are attached by taking
into account both the count statistics and the typical rms background
fluctuations occurring on the same scale as the aperture. The latter
quantity is computed by evaluating the surface brightness in a number
of non-overlapping apertures spread throughout the frame and located
at least 400 pixels from the centre of the galaxy.

\section{Results}
\begin{figure}
\includegraphics[width=8truecm]{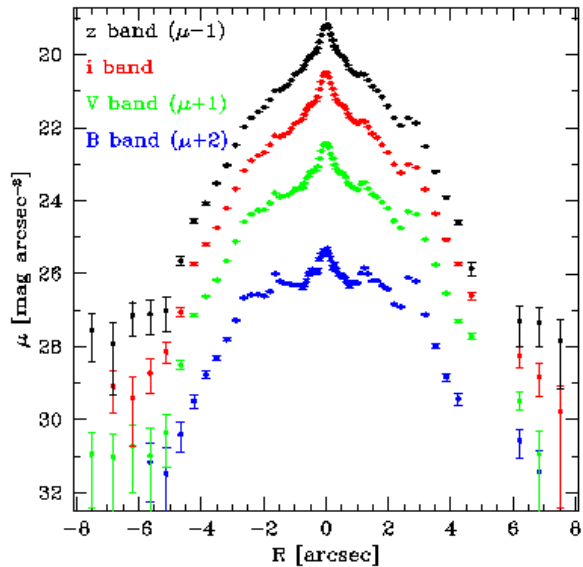}
\caption{Major axis surface brightness profiles. Negative $R$ are used for the
eastern side, positive for the western. Error bars are computed as explained in the text.
Only points with S/N$>$1 are plotted. The profiles for the different bands are 
coded in different colours and offset in
order to avoid confusion, as indicated in the legend.}\label{majorprof}
\end{figure}

\begin{figure}
\includegraphics[width=8truecm]{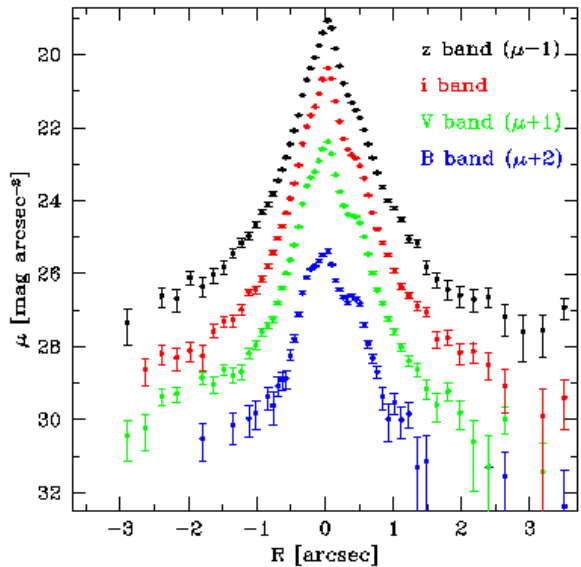}
\caption{Same as Fig.\ref{majorprof} for the minor axis. Negative $R$ are used for the
southern side, positive for the northern.}\label{minorprof}
\end{figure}

In Fig. \ref{majorprof} and \ref{minorprof}, we show the surface
brightness profiles measured along the major and minor axes of the
galaxy respectively. As described in the legend, the different
coloured symbols correspond to the four bands.  The
asymmetry seen in the inner regions of both profiles attests to the
presence of spiral structure and dust lanes in these parts.  We
include these regions for completeness, but note their interpretation
is complicated; our main focus is the behaviour of the profiles at
very large radius.


The major axis profile is extracted using wedges aligned with the $+x$
and $-x$-axes and with a small opening angle of 10$\degr$ (see
Fig. \ref{majorprof}). While a warp is present in the outer parts of the disk,
our extraction method involves summing over many pixels at
large radius and is thus not significantly affected by the deviation from the midplane. 
 In all four bands, the profile shows a
profound change in slope at $\sim 3$ arcsec. By fitting an exponential
function to the surface brightness profile of the bright inner disc 
(corresponding to the region where
highly structured spiral arms and dust lanes are visible in
Fig.\ref{picture}), we find that the projected scalelength
varies with wavelength from $r_{\mathrm{inner},B}=4.2$ arcsec to
$r_{\mathrm{inner},z}=1.3$ arcsec. The steeper outer profile is very
similar in all bands, with exponential scalelength of $\sim0.55$
arcsec. Beyond 5 arcsec an excess with respect to the exponential
profile is seen in the three reddest bands at the level of $\mu
\sim29-30$~mag arcsec$^{-2}$.

The minor axis profile is extracted using wedges aligned with the the
$+y$ and $-y$-axes and with a wider opening angle of 45$\degr$ in
order to ensure adequate S/N far from the disc plane (see
Fig. \ref{minorprof}).  On the southern side of the galaxy, the inner
profile is more regular and declines as a smooth exponential until
approximately 0.5 arcsec.  Beyond 1 arcsec, a significant power-law
excess is apparent (on both sides of the plane) in all four bands (see
Fig. \ref{minorprof}).  The slope of this power-law component is
measured to be $\sim 2.6$ in the three reddest bands; we expect the
effect of dust (as well as emission from an inclined disc) 
 to be minimal at these heights ($\gtrsim 5$~kpc) from
the midplane.  

We have checked that the point spread function (PSF) does not have a
significant effect on our derived profiles by studying the surface
brightness profile of a relatively bright, unsaturated star
($m_V\sim20.5$ mag) which falls within the UDF.  We find that the
surface brightness drops by $\gtrsim 10 (12)$ mag arcsec$^{-2}$ from
the centre to 1 (2) arcsec in all bands.  As a result, scattered light
from the central PSF should be well below our measurement
uncertainties.

In order to estimate the isophotal ellipticity of the power-law
component, we measure the surface brightness as a function of radius
in narrow wedges (13$\degr$ wide) placed at different position
angles. By comparing surface brightnesses as a function of position angle out
to a radius of 5.5 arcsec, we derive an axial ratio of $b/a\sim0.45$
in the $i$-band at $\mu_{i}\sim29$.  The outermost isophotes  thus appear 
significantly rounder than the inner (disc-dominated) isophotes. Our measurement represents a lower limit on
the actual flattening however, since low-level disc emission may still
contaminate the isophotes at 0 and 180$\degr$. Indeed, if we exclude
the sectors which lie within 25$\degr$ of the major axis, the
ellipticity $b/a$ of the faint outer component increases to 0.6.

\begin{figure}
\includegraphics[width=8truecm]{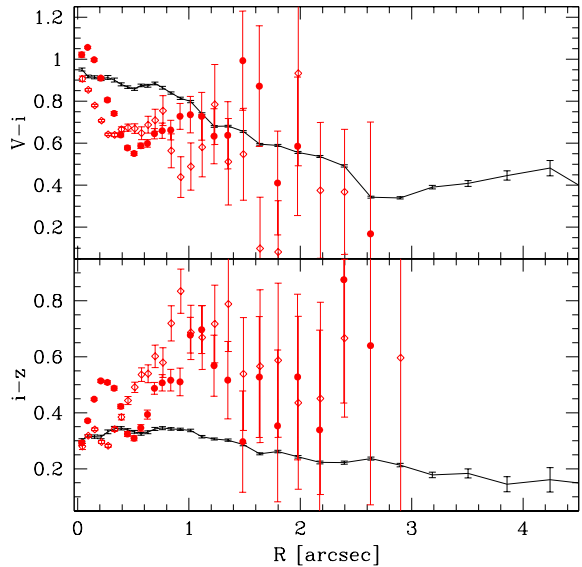}
\caption{Colour profiles: $V-i$ (top), $i-z$ (bottom).
Black solid lines represent the major axis and red points represent the minor axis (filled circles
for the northern side, open diamonds for the southern). See text for details.}\label{colours}
\end{figure}

In Fig. \ref{colours} we plot the $V-i$ and $i-z$ colours along the
major (black solid line) and minor axes (red points) as obtained from
combining the surface brightness profiles shown in Figures
\ref{majorprof} and \ref{minorprof}.  The major axis profile
represents the average of the eastern and western axes in regions
where S/N$>2$.  Error bars are attached using the standard propagation
of the errors on the surface brightness measurements.  The profiles
along the major axis display a clear blue gradient in $V-i$ within
$\sim2.5$~arcsec, where the bright disc dominates. Beyond $\sim3$
arcsec, where the exponential slope significantly steepens, the
gradient in this colour is inverted.  The $i-z$ profile is
significantly flatter, slowly varying from 0.3--0.35 in the inner 1
arcsec, to 0.15 at 4 arcsec.\\
For the minor axis, the north and south sides are shown
separately as filled circles and open diamonds respectively.  Along
the minor axis we observe a clear blue gradient in $V-i$ within
$\sim0.5$ arcsec, while the $i-z$ profile in this region is very
irregular and likely reflects both internal structure as well as dust
within the disc.  The transition between the disc and power-law regions  
is characterised by a strong red
gradient in $i-z$, from 0.4 at 0.5 arcsec to 0.7 at 1 arcsec
(considering the average of the northern and southern profiles).  In
the region dominated by the power-law excess ($R\gtrsim1.5$ arcsec),
the colour measurements are rather noisy due to the very low surface
brightness, but still there is evidence for $V-I\sim0.6-0.7$ and
$i-z\sim0.5\pm0.1$.

We also estimate the total magnitude of the galaxy, of the disc
alone, and of the galaxy excluding the disc in each band 
(see Table \ref{magtab}). The total magnitude is
integrated in an ellipse with $b/a=0.6$, $a=8$ arcsec, while for the
disc alone we adopt an ellipse with $b/a=0.2$, $a=4.8$ arcsec. The
``Total$-$Disc'' magnitude is integrated in the complementary region. 
Typical uncertainties are of order of 0.05 for the ``Total$-$Disc'',
essentially given by count statistics and background uncertainty.
In the three reddest bands the contribution of the non-disc component
ranges from 4.3 to 5.6 per cent of the total light and displays colours which are
significantly redder than those of the disc, namely $V-i=0.73\pm0.10$
vs. 0.64 and $i-z=0.52\pm0.10$ vs. 0.33.  Reddening due to dust is unlikely to 
affect the colours at these radii and, in any case, any realistic distribution
of dust would lead to bluer colours with increasing distance from
the disc instead of redder ones.
%bluer colours at large radii instead of the redder colours
%observed. 

\begin{table*}
\caption{Integrated photometry and colours: the total and disc fluxes are integrated
in elliptical apertures, as described in the text. For the ``Disc'' and ``Total$-$Disc''
components, the percentage over the total flux are given as well.}\label{magtab}
\begin{tabular}{l | rr | rr | rr | r | r }
\hline
&
\multicolumn{2}{c |}{$V$}&
\multicolumn{2}{c |}{$i$}&\multicolumn{2}{c |}{$z$}&
$V-i$&$i-z$\\
&mag&\%&mag&\%&mag&\%&
mag&mag\\
\hline
Total     &21.14&          &20.50&          &20.16&      &0.64&0.34\\
Disc      &21.18&95.7\%&20.55&95.4\%&20.23&94.4\%&0.64&0.33\\
Total$-$Disc&24.54& 4.3\%&23.82& 4.6\%&23.30& 5.6\%&0.73&0.52\\
\hline
\end{tabular}
\end{table*}

\section{Discussion}

Our analysis of the UDF galaxy, COMBO-17 \#31611, has yielded the
detection of a very low surface brightness structural component in
addition to the bright disc.  The main evidence for this component is:
{\it i)} excess emission with respect to an exponential profile
detected to $3$ arcsec ($\sim$14 kpc) from the disc plane at the level
of $\mu_{V,i}\sim29$ mag~arcsec$^{-2}$; {\it ii)} the isophotal shape
is centrally-concentrated and 
becomes rounder with increasing radius, reaching $b/a\sim0.6$ at the
faintest measured isophote; {\it iii)} colours far from the plane that
are distinct from the disc, \ie $i-z\sim0.5$ and $V-i\sim0.7$ which
are 0.2-0.3 and 0.1 mag redder than the disc respectively.  Based on
these properties, we attribute this component to the stellar halo of
the galaxy, making it the most distant detection of a stellar halo yet
known.  We note the striking resemblance between the halo properties
derived here and those of the ``mean'' halo detected in Z04's SDSS
stacking analysis. As discussed in Z04, these observations are also
consistent with the few extant observations of individual stellar
haloes.

The surface brightness profile of the halo is consistent with a
power-law falling as $I\propto R^{-2.6}$. This slope must be
considered an upper limit since if there is contaminating outer disc emission
within our aperture, it will steepen the observed profiles even beyond 1
arcsec. An $R^{1/4}$ law provides a poor fit to the data, which
display excess emission at the faintest surface brightness levels
compared to this profile. Less concentrated S\'ersic profiles, which
are more appropriate for the small bulges of late-type galaxies,
provide even worse fits, thus ruling out the bulge as responsible
for the measured emission.
The fractional contribution of the halo to
the total galaxy light is around 5 per cent, but this should be
considered only approximate since we neglect disc contamination in the
region where the halo light is integrated and include the inner halo
as part of the disc light. Furthermore, we have no handle on internal
extinction which could be significant in the inner regions of the
galaxy, where the disc contribution is calculated.

In order to conduct a direct comparison between the surface brightness
and colours of this distant stellar halo and those of local galaxies,
corrections need to be applied for bandpass shifting and surface
brightness dimming. Given the photometric redshift of 0.322, the four
observed bands, $BViz$ correspond rather well to the SDSS $ugri$ bands
at redshift zero. Assuming a 9~Gyr, $Z=0.4 Z_\odot$ single stellar
population (SSP), we use the
\cite{BC03} (hereafter BC03) models to calculate that  observed surface 
brightnesses in $Viz$
translate into rest-frame 
values of $\sim1.1$ mag~arcsec$^{-2}$ brighter in $gri$.  Our
measurements of $\mu_{V,i,z}\sim29-30$ mag~arcsec$^{-2}$ thus
correspond to rest-frame equivalents of $\mu_{g,r,i}\sim28-29$
mag~arcsec$^{-2}$.  This is precisely the surface brightness range in which
a power-law halo component starts to dominate the minor axis 
profile in Z04's SDSS stacking analysis and in which direct detections
of individual haloes have been made in nearby galaxies.\\
\indent
We proceed to use the BC03 models to interpret the colours of the halo
emission in terms of stellar populations.  First, we compare the
observed colours with predictions for different SSPs with ages of
2--13~Gyr and metallicities of 0.005--2.5~$Z_\odot$ adopting a
\cite{Chabrier03} Initial Mass Function (IMF).  We find none of these
models can simultaneously reproduce the intermediate $V-i$ and the
very red $i-z$ colours. Unless a super-solar metallicity is assumed
($Z=2.5 Z_\odot$), the measured $i-z\sim0.5$ is at least 0.15 mag
redder than the models and thus inconsistent at more than the 1.5
$\sigma$ level. Such an anomaly is reminiscent of the finding of a
correspondingly high $r-i\sim0.6$ by Z04, despite the intermediate
$g-r\sim0.65$ that could be  accounted for by an old 
population with roughly solar metal enrichment.\\
\indent
Since we have been unable to produce a model that provides a good fit
to the observed colours at $z=0.32$, we adopt the following strategy
to derive rest-frame colours that can be compared to haloes in the
local universe.  We adopt a fiducial reference model of an SSP of age
9~Gyr at the redshift of the galaxy, with $Z=0.4~Z_\odot$, and a
Chabrier IMF (BC03); this model minimises the absolute deviations from
the observed colours.  We then evolve the spectrum to $z=0$ and derive
the rest-frame colours that would be observed through SDSS $gri$
passbands.  These colours can then be directly compared to the
measurements of Z04 for the $``$mean'' halo in their SDSS stack.  Not
unexpectedly, the evolved colours do not provide a good match to
Z04's observations but interestingly the offset between the model and
observations is of the same magnitude and sense at $z=0$ as at
$z=0.32$. Specifically, while the model predicts a blue colour ($V-i$
or SDSS $g-r$) that is consistent or just slightly redder than the
observations, the predicted red colour ($i-z$ or SDSS $r-i$) is too
blue by $\sim0.2$ mag.\\
This kind of colour anomaly, \ie too much flux
in the spectral region around 7500\AA~with respect to shorter optical
wavelengths, was also seen in the stellar halo of the nearby system
NGC~5907 \cite{lequeux_etal98}.  An attempt to directly resolve the
population of metal-rich giant stars that would give rise to these red
colours was unsuccessful, leading to the suggestion that the stellar
halo in NGC~5907 formed with a non-standard IMF that is dominated by
$M\lesssim 0.2M_\odot$ stars
\citep{zepf_etal00}.   While such a truncation could account for
the lack of halo giant stars, it is unclear whether it could also
explain the  colours observed here, \ie $g-r$ typical of G-K stars yet
$r-i$ typical of M stars.  Another possibility is that the red
colour is contaminated by ionised gas emission.  Deep $H\alpha$
imagery of nearby galaxies often reveals faint extraplanar emission
\citep{ferg96,mill03}, however little is known about the extent of
these ionised ``haloes'' at very faint flux levels.  Further
investigation is clearly required in order to understand the origin of
the peculiar halo colours, which have now been measured in three
independent studies.

As a final remark, we note that the surface brightness profile of the disc
can be fitted by an exponential function that changes scalelength abruptly
at 3~arcsec~$\simeq14~\mathrm{kpc}$ (corresponding to $\sim 2.4$
inner scalelengths).   This behaviour is consistent with
deep imaging studies of both edge-on and face-on discs in the local
universe \citep{kregel_etal02, pohlen02} and likely reflects a
truncation of the bright inner disc.  That this radius also represents
the point where the major axis colour gradient reverses and starts
to redden is intriguing and suggests that the stellar populations
at the far extremity of the disc are of significant age.\\

\thanks{We thank the referee, Michael Pohlen, for useful comments. 
The research of AMNF has been supported by a Marie
Curie Fellowship of the European Community under contract number HPMF-CT-2002-01758.}

 \label{lastpage}

\end{document}